# Oriented External Electric Field Modulated Second-Order Nonlinear Optical Response and Visible Transparency in Hexalithiobenzene

Ambrish Kumar Srivastava

*Department of Physics, Deen Dayal Upadhyaya Gorakhpur University, Civil Lines, Gorakhpur 273009, India*

E-mail: ambrishphysics@gmail.com; aks.ddugu@gmail.com

**Abstract**

Oriented external electric fields (OEEFs) offer a unique opportunity to tune certain activity of molecules by orienting the alignment of the electric field along the specific axis. The second-order NLO response of hexalithiobenzene ($C_6Li_6$) is very poor due to its first mean hyperpolarizability of 0.5 a.u. Therefore, we have analyzed the effect of OEEFs on the structural, electronic properties and NLO response of $C_6Li_6$ using a density functional approach. We notice that the structure of the $C_6Li_6$ molecule remains planar, with the slight change in C-C and C-Li bond lengths but their stability is increased under the effect of OEEFs. By applying OEEFs, the conductivity or reactivity of $C_6Li_6$ is increased as their HOMO-LUMO energy gap is decreased. Furthermore, $C_6Li_6$ attains finite dipole moment in the presence of OEEF, which increases linearly as the OEEF increases. More interestingly, the first mean hyperpolarizability of $C_6Li_6$ is significantly enhanced, becoming as high as 3.4 $\times 10^4$ a.u. for OEEF = 50 $\times 10^{-4}$ a.u. This suggests the OEEF as an effective way to enhance the second-order NLO responses, leading to the design of potential NLO materials. Nevertheless, the visible transparency of $C_6Li_6$ with and without OEEF may suggest its possible applications in optical devices.

**Keywords:** Hexalithiobenzene; External electric field; NLO response; Visible transparency; DFT calculations.

## 1. Introduction

The nonlinear optical (NLO) behavior of materials led to various applications in optoelectronics, electro-optics, rechargeable batteries, sensors, detectors, catalysis, *etc.* [1, 2]. In general, the NLO response of systems is enhanced either by electron push-pull mechanism having appropriate electron donor and acceptor groups linked through π conjugation [3] or by metal-ligand frameworks in organometallic compounds [4]. However, Chen *et al.* [5] and Li *et al.* [6] revealed that the introduction of excess electrons [7] in a system is probably the most effective strategy for the enhancement of NLO responses. For instance, the mean hyperpolarizability of $(H_2O)_3$ cluster with the excess electron is $10^6$ times higher than that of equivalent molecular cluster lacking excess electron [5, 7]. In the past few years, various NLO materials have been synthesized including barium borate ($BaB_2O_4$) crystal [8], alkali-metal fluorooxoborates ($MB_4O_6F$; $M$ =K, Rb, Cs) [9], $Ca_3(O_3C_3N_3)_2$ crystal [10], *etc*. Recently, 2D carbon materials have caught attention and the NLO response of various alkali atom or cluster doped graphene (GE), graphyne (GY), graphdiyne (GDY), etc. have been reported [11, 12].

It is now widely accepted that the oriented external electric fields (OEEFs) play an important role in the chemistry of materials. OEEFs have been found to provide effective control over reactivity and selectivity of chemical reactions [13-16] by reorganizing the electron distribution of molecules. Shaik *et al.* [17] suggested that the OEEFs possess the potential of becoming smart catalysts or inhibitors of non-redox reactions and as controllers of reaction mechanisms. OEEFs have been also investigated to facilitate spin-crossover transitions [18], spin-polarized conductivity [19], carbon nanotube growth [20], and the activation of compounds such as methane [21] or carbon dioxide [22]. More recently, the OEEF has been utilized to tune the electronic properties and NLO responses of some superatom compounds, including (NLi$_4$)(BF$_4$) and (BLi$_6$)X (X = BeF$_3$ and BF$_4$) [23]. This prompted us to study the effect of the OEEF on the properties of hexalithiobenzene (C$_6$Li$_6$).

C$_6$Li$_6$, a fully lithiated analogue of benzene (C$_6$H$_6$), was synthesized in 1978 by Shimp *et al.* [24] and isolated in 1992 by Baran *et al.* [25]. It has a planar star-like structure, which has been previously studied by several researchers [26-29]. Apart from its structural beauty, its low ionization energy feature [30] and application in hydrogen storage have been also reported [31, 32]. Recently, C$_6$Li$_6$ has been proposed for the activation and enhanced storage of CO$_2$ molecules [33]. Raptis *et al.* [34] have recognized the exceptionally high second-order hyperpolarizability of the C$_6$Li$_6$ molecule, responsible for the third harmonic generation. However, the second-order NLO response of C$_6$Li$_6$ is expected to be vanishingly small due to its planar and symmetric structure. A recent study [23] suggested that the NLO response can be significantly enhanced by applying the OEEFs on some superatom compounds. Note, however, that these compounds are ionic, having significant NLO response even in the absence of OEEF. This leads us to know whether OEEF can be utilized to modulate the electronic properties and in particular, NLO responses of C$_6$Li$_6$. In order to explore this, we have studied the electronic properties and NLO responses of C$_6$Li$_6$ by varying the OEEF from 10 to 50 ×10$^{-4}$ a.u. using density functional theory (DFT) based approach. In addition, we have focused on the stability of C$_6$Li$_6$ under external electric field as well as its visible transparency.

2. **Computational details**

All DFT calculations on C$_6$Li$_5$ were performed at the B3LYP method [35, 36] using a 6-311+G(d) basis set in the Gaussian 09 program [37]. To prove the reliability of our method, we have performed some test calculations on benzene (in the absence of the experimental data on C$_6$Li$_6$). Our B3LYP computed C-C bond lengths (1.395 Å) and VIE (9.28 eV) of C$_6$H$_6$ are in good agreement with the experimental bond lengths of 1.399 Å measured by infrared spectroscopy [38] and ionization energy of 9.24 eV [39], respectively. The OEEF was applied along the *z*-axis specifying the *field = z + n* keyword with $n$ = 0, 10, 20, 30, 40 and 50. The geometry optimization was followed by vibrational frequency calculations to ensure that the optimized structures correspond to true minima in the potential energy surface. The partial atomic charges were computed by a natural bond orbital (NBO) scheme, as implemented in the Gaussian 09 program.

The NLO parameters such as dipole moment, polarizability and first-order hyperpolarizability were obtained at CAM-B3LYP [40], recommended for long-range interactions and NLO properties. The total energy ($E$) of a molecular system in the presence of an external electric field can be expressed as [41],

$$E = E^0 - \mu_i F_i - \frac{1}{2}\alpha_{ij} F_i F_j - \frac{1}{6}\beta_{ijk} F_i F_j F_k - \cdots \quad (1)$$

where $E^0$ is the total energy in the absence of the electric field, and $F_i$, $\mu_i$, $\alpha_{ij}$ and $\beta_{ijk}$ represent the components of the electric field, dipole moment, polarizability and hyperpolarizability, respectively along with the directions specified by subscripts, *i*, *j* and *k* = *x*, *y* and *z*. Using the finite-field approach [42], the mean polarizability ($\alpha_0$) and first-order mean hyperpolarizability ($\beta_0$) can be obtained as follows:

$$\alpha_0 = \frac{1}{3}(\alpha_{XX} + \alpha_{YY} + \alpha_{ZZ}) \quad (2)$$

$$\beta_0 = (\beta_X^2 + \beta_Y^2 + \beta_Z^2)^{\frac{1}{2}}; \quad \beta_i = \frac{3}{5}(\beta_{iii} + \beta_{ijj} + \beta_{ikk}) \quad (3)$$

### 3. Results and discussion

The structure of the planar $C_6Li_6$ molecule is displayed in Fig. 1 as mentioned earlier. Although the structure is symmetric, a vanishingly small dipole moment of ~0.002 D (1 D ≈ 3.3356 ×10$^{-30}$ C.m) is obtained for the equilibrium structure of $C_6Li_6$ without considering any symmetry constraints during optimization. The direction of this dipole moment ($\mu_Z$) is also depicted in Fig. 1. Conventionally, the direction of $\mu_Z$ is taken from negative to positive charge regions. Therefore, we have applied OEEF varying from $10 \times 10^{-4}$ a.u. to $50 \times 10^{-4}$ a.u., where 1 a.u. = $5.142 \times 10^{11}$ V/m, in the direction opposite to that of $\mu_Z$. With no OEEF, $C_6Li_6$ has equal C-C bond lengths of 1.420 Å and C-Li bond lengths of 1.914 Å. As OEEF increases, this bond-length equalization is no longer preserved but the planarity of the ring remains unperturbed.

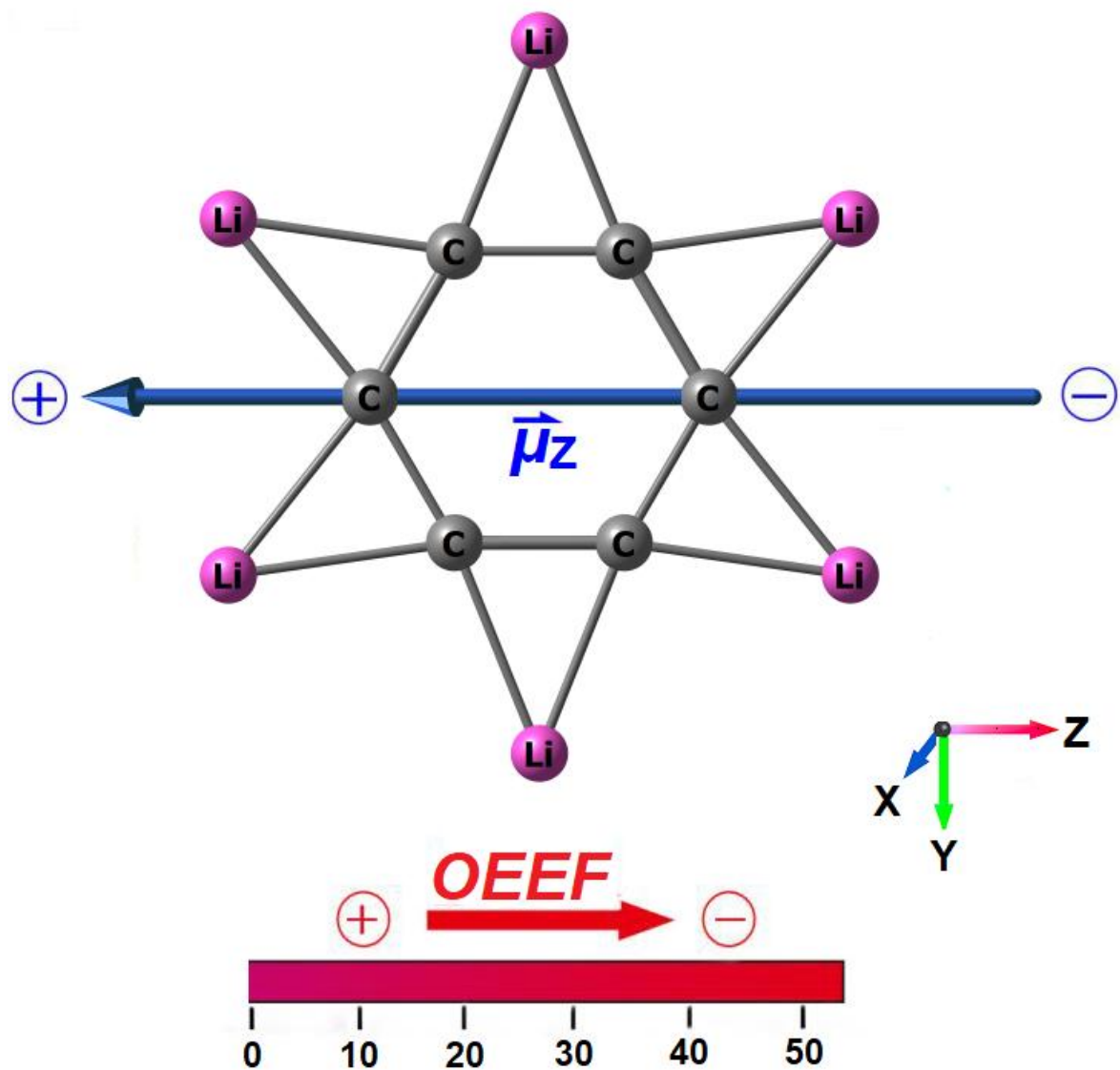

Fig. 1. Model structure of $C_6Li_6$ with the directions of dipole moment and oriented external electric field (OEEF) to be applied (in units of $10^{-4}$ a.u.).

The structural evolution of the $C_6Li_6$ molecule with the increase in the magnitude of OEEF is displayed in Fig. 2. For OEEF = 10 ×10$^{-4}$ a.u., there is no change in the bond-lengths of $C_6Li_6$. For OEEF = 20 ×10$^{-4}$ a.u., the C-Li bond lengths range between 1.911-1.921 Å, however, C-C bond lengths are not changed. For OEEF = 30 and 40 ×10$^{-4}$ a.u., the C-C bond-lengths tend to change and C-Li bond lengths vary between 1.909-1.925 Å and 1.909-1.930 Å, respectively. For OEEF = 50 ×10$^{-4}$ a.u., the C-C bond lengths lie in the range 1.417-1.422 Å and C-Li bond lengths become 1.909-1.935 Å. These geometrical features are closely associated with the charge distribution discussed below.

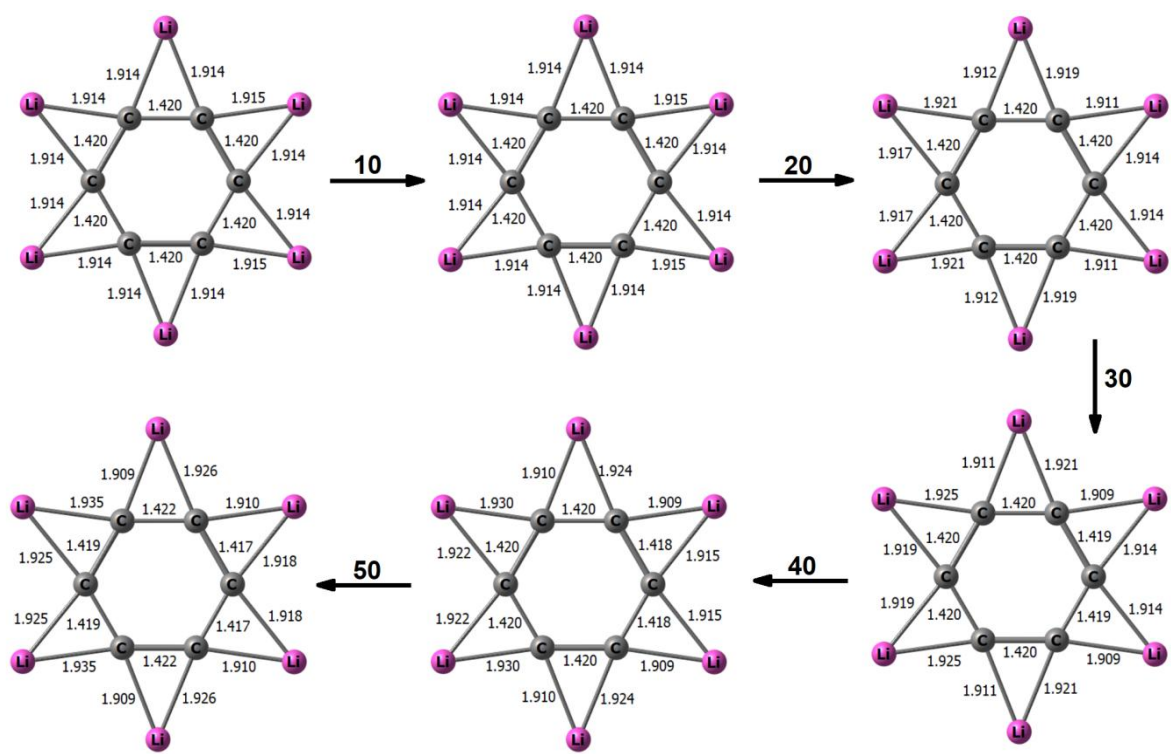

Fig. 2. Structural evolution of $C_6Li_6$ molecule under the applied OEEF ((in units of 10$^{-4}$ a.u.).

The total energy of a $C_6Li_6$ molecule is listed in Table 1. The total energy without OEEF ($E^0$) is calculated to be -273.772644 Hartree. One can see that there is a decrease in energy ($E$) with an increase in the OEEF value. Furthermore, the difference between $E$ and $E^0$ also increases rapidly with the increase in the OEEF. For instance, the $E$ is lower than $E^0$ (as expected) by 0.000791 Hartree (~0.02 eV) for OEEF = 20 ×10$^{-4}$ a.u. and 0.005229 Hartree (~0.14 eV) for OEEF = 50 ×10$^{-4}$ a.u. This may indicate that the electronic stability of the $C_6Li_6$ molecule increases with the increase in OEEF. According to Moreno *et al.* [43], the stability of the planar $C_6Li_6$ molecule is lower than the structure with three $C_2^{2-}$ fragments strongly aggregated through lithium bridges. Thus, our study provides a way to increase the stability of a molecule by applying external electric fields.

Table 1. B3LYP/6-311+G(d) calculated total energy (*E*), HOMO energy ($E_{HOMO}$), LUMO energy ($E_{LUMO}$), HOMO-LUMO gap ($E_{gap}$), NBO charge on Li ($Q_{Li}$) and vertical ionization energy (VIE) of $C_6Li_6$ complexes for oriented external electric fields (OEEFs) applied.

| OEEF (10$^{-4}$ a.u.) | *E* (Hartree) | $E_{HOMO}$ (eV) | $E_{LUMO}$ (eV) | $E_{gap}$ (eV) | $Q_{Li}$* (*e*) | VIE (eV) |
|---|---|---|---|---|---|---|
| 0 | -273.772644 | -3.065 | -1.171 | 1.894 | 0.618 | 4.482 |
| 10 | -273.772834 | -3.065 | -1.175 | 1.891 | 0.603-0.633 (0.618) | 4.484 |
| 20 | -273.773435 | -3.074 | -1.334 | 1.741 | 0.586-0.647 (0.618) | 4.484 |
| 30 | -273.774441 | -3.088 | -1.511 | 1.577 | 0.568-0.660 (0.617) | 4.489 |
| 40 | -273.775890 | -3.107 | -1.807 | 1.299 | 0.548-0.673 (0.615) | 4.496 |
| 50 | -273.777873 | -3.144 | -2.284 | 0.859 | 0.519-0.686 (0.613) | 4.495 |

*Average values, wherever required, are given in parentheses.

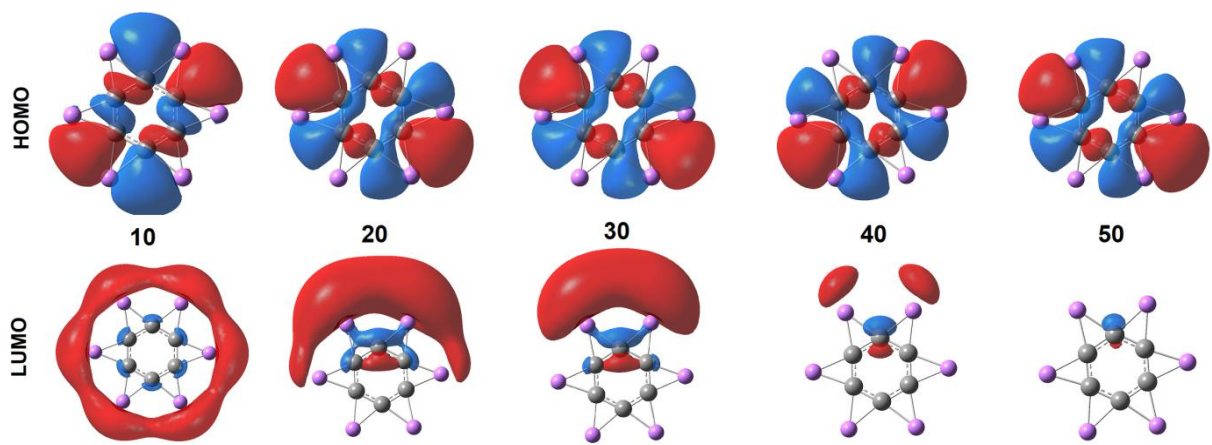

Fig. 3. The frontier orbitals' surfaces of $C_6Li_6$ molecule under the applied OEEF ((in units of 10$^{-4}$ a.u.).

The frontier orbitals, the highest occupied molecular orbital (HOMO) and lowest unoccupied molecular orbital (LUMO), serve as an important indicator to predict the chemical reactivity of molecules and their interactions with other species. The HOMO and LUMO surfaces of $C_6Li_6$ are plotted in Fig. 3 for different OEEF values. For OEEF = $10 \times 10^{-4}$ a.u., one can see that the HOMO of $C_6Li_6$ is composed of the atomic orbitals of C and Li atoms. However, the LUMO is mainly contributed by the atomic orbital of Li atoms, including a small contribution from C-atoms. With the increase in the OEEF, the composition of HOMO does not change, but the number of atomic orbitals contributing to the LUMO decreases. We will see later that this results in the change in the LUMO energy level, leading to a smaller HOMO-LUMO gap. The energies of HOMO ($E_{HOMO}$), LUMO ($E_{LUMO}$) and HOMO-LUMO gap ($E_{gap}$) are also listed in Table 1. The $E_{gap}$ of a molecular system is considered to be an important electronic property, which can represent its ability to participate in chemical reactions to some extent. Moreover, the $E_{gap}$ is also closely related to the band-gap in solids, which is an important parameter to quantify the conductivity of materials. From Table 1, the $E_{gap}$ of $C_6Li_6$ without the external electric field is as large as 1.89 eV, exhibiting its semiconducting properties. The dependence of $E_{gap}$ on OEEF values is plotted in Fig. 4. It is evident that the $E_{gap}$ of $C_6Li_6$ gradually reduces from 1.89 to 0.84 eV with the increasing electric field from 0 to $50 \times 10^{-4}$ a.u. This variation of the HOMO-LUMO gap suggests that the conductivity of $C_6Li_6$ can be enhanced so that this may exhibit metallic properties. To understand the origin of the change of $E_{gap}$, we analyze the variations of $E_{HOMO}$ and $E_{LUMO}$ as a function of OEEF. From Fig. 4, one can observe that the increasing electric field strength leads to a decrease in the LUMO level. Note that the HOMO level remains almost constant and decreases only marginally under the same OEEFs. Therefore, the HOMO-LUMO gap of $C_6Li_6$ mainly depends on the evolution of the LUMO level under the varying OEEFs.

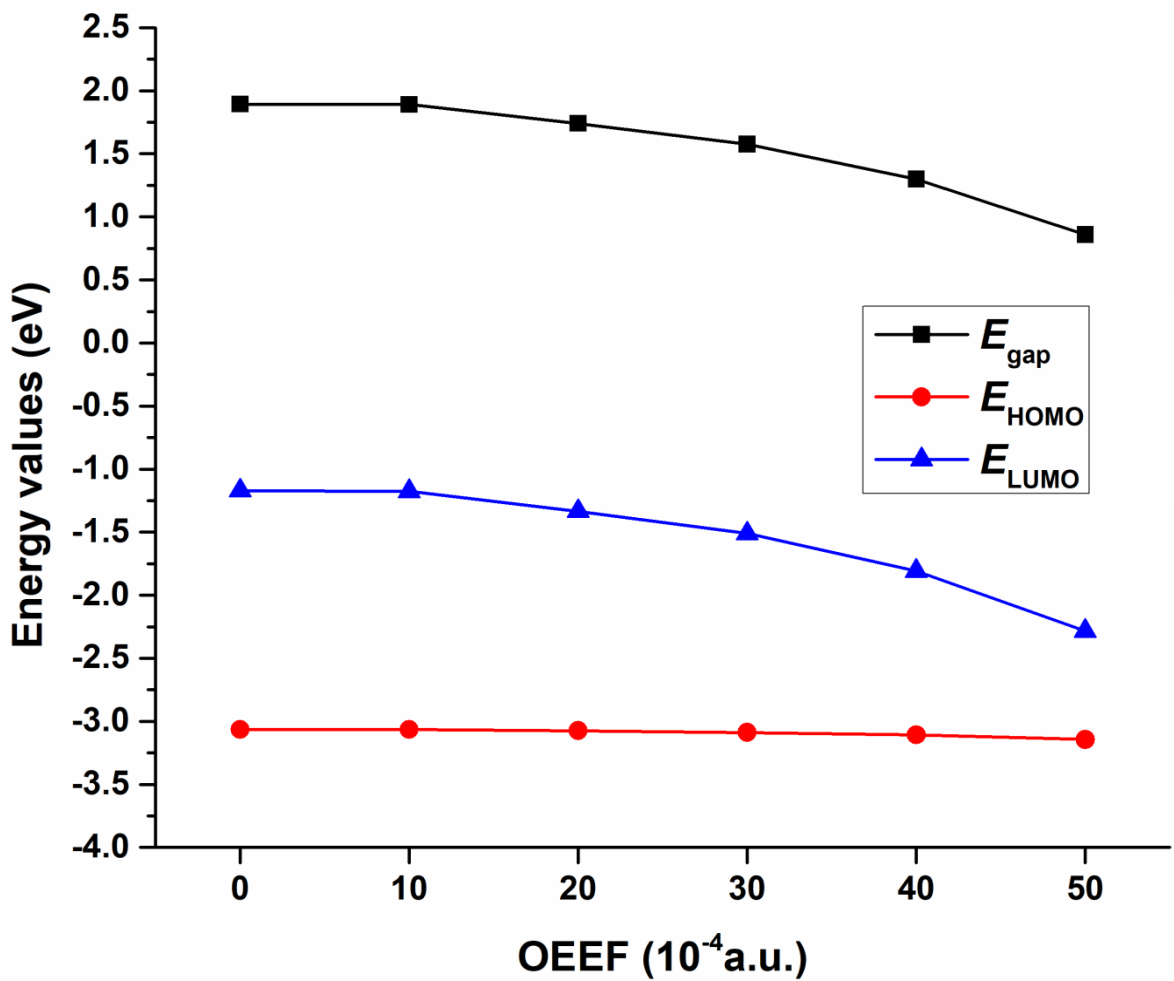

Fig. 4. The variation of HOMO energy ($E_{HOMO}$), LUMO energy ($E_{LUMO}$), HOMO-LUMO gap ($E_{gap}$) of $C_6Li_6$ molecule as a function of oriented external electric field (OEEF).

The vertical ionization potential (VIE) can be considered as an important parameter for estimating the electronic stability of the molecular system. The VIE values of $C_6Li_6$ for various OEEFs have been calculated and are also listed in Table 1. It is observed that the VIE values are almost constant or slightly increased by varying OEEF from $10 \times 10^{-4}$ to $50 \times 10^{-4}$ a.u. The IE can be closely related to $E_{HOMO}$ according to Koopmans' theorem [44]. Thus, the trend of VIEs may be expected due to the fact that the $E_{HOMO}$ values of $C_6Li_6$ do not vary significantly with increasing OEEF (see Table 1 and Fig. 4). These constant or slightly enhanced VIEs support that the electronic stability of $C_6Li_6$ is gradually increased by applying external electric fields. This fact is also consistent with the total energy of $C_6Li_6$ discussed earlier. To analyze the charge distribution of $C_6Li_6$ under the effect of the external field, we have listed natural charges on Li atoms in Table 1. There is a charge transfer from terminal Li atoms to ring C atoms so that each Li possesses an equal positive charge ($Q_{Li}$) of 0.618$e$. This leads to the bond-length equalization and negligible dipole moment as mentioned earlier. By applying the external field, the charge distribution remains no longer isotropic. For instance, $Q_{Li}$ lies in the ranges 0.603-0.633$e$ and 0.586-0.647$e$ for OEEF = 10 and 20 $\times 10^{-4}$ a.u., respectively, although average $Q_{Li}$ remains 0.618$e$. This corresponds to the variation in C-Li bond lengths and leads to an increase in the dipole moment, as listed in Table 2. With the increase in OEEF from 30 to 50 $\times 10^{-4}$ a.u., the average $Q_{Li}$ varies from 0.617$e$ to 0.613$e$. This results in the change of C-Li as well as C-C bond lengths and further increases the dipole moment.

Table 2. CAM-B3LYP/6-311+G(d) computed dipole moment ($\mu_Z$), mean polarizability ($\alpha_0$), first-order hyperpolarizabilities ($\beta_i$; $i$= X, Y and Z) and mean hyperpolarizabity ($\beta_0$) of $C_6Li_6$ complexes for oriented external electric fields (OEEFs) applied. TD-DFT//CAM-B3LYP/6-311+G(d) calculated absorption wavelength ($\lambda_{max}$) is also listed.

| OEEF ($10^{-4}$a.u.) | $\mu_Z$ (Debye) | $\alpha_0$ (a.u.) | $\beta_X$ (a.u.) | $\beta_Y$ (a.u.) | $\beta_Z$ (a.u.) | $\beta_0$ (a.u.) | $\lambda_{max}$ (nm) |
|---|---|---|---|---|---|---|---|
| 0 | 0.002 | 241.5 | 0 | 0 | 0.5 | 0.5 | 572.7 |
| 10 | 0.304 | 242.2 | 1.8 | 5.4 | 2574.0 | 2574.0 | 569.8 |
| 20 | 0.643 | 244.9 | 6.6 | 12.6 | 5439.0 | 5439.0 | 563.1 |
| 30 | 0.975 | 249.4 | 13.8 | 25.8 | 9272.4 | 9272.5 | 560.9 |
| 40 | 1.321 | 257.0 | 34.8 | 67.2 | 15649.8 | 15650.0 | 528.9 |
| 50 | 1.700 | 272.2 | 147.6 | 406.8 | 33867.6 | 33870.4 | 634.8 |

Since $\mu_X$ and $\mu_Y$ components are zero, only the $\mu_Z$ component contributes to the dipole moment of $C_6Li_6$. From Table 2 and Fig. 5, one can note that the $\mu_Z$ increases monotonically by increasing the external field. Table 2 also lists mean polarizabilities ($\alpha_0$) of $C_6Li_6$ as a function of OEEF values, which are also plotted in Fig. 5 for visual indication. The mean polarizabilities of $C_6Li_6$ gradually increase with the increase in the OEEFs, lying in the range 241.5−272.2 a.u. According to the hard-soft acid-base (HSAB) principle [45], the molecules with smaller $E_{gap}$ are softer and hence, more polarizable. From Fig. 4 and Fig. 5, one can observe that the increase in polarizability is in accordance with the decrease in the HOMO-LUMO gap.

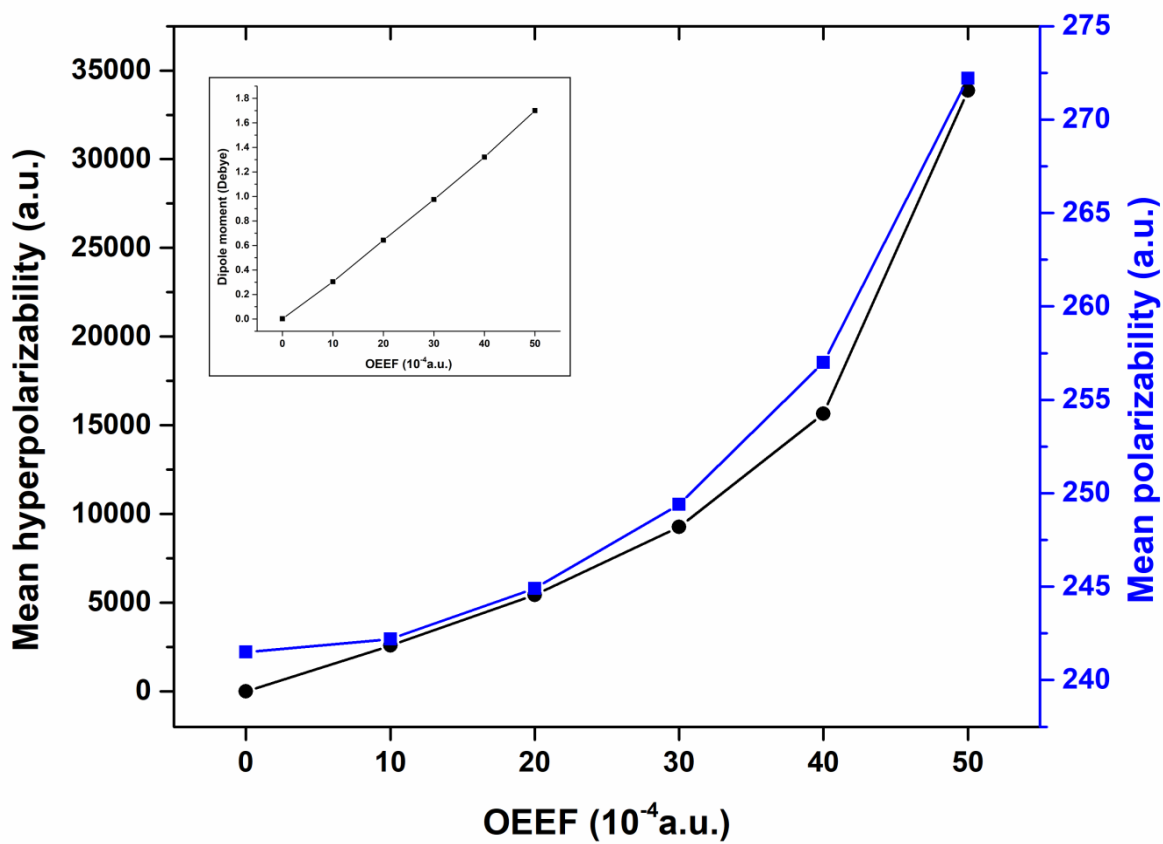

Fig. 5. The variation of mean polarizability and mean hyperpolarizability of $C_6Li_6$ molecule as a function of oriented external electric field (OEEF). The variation of dipole moment has been also displayed (inset).

The nonlinear optical (NLO) behaviour of $C_6Li_6$ molecule on applying the external field has been explored by computing the first-order hyperpolarizabilities ($\beta_X$, $\beta_Y$ and $\beta_Z$), which are also listed in Table 2. The mean hyperpolarizability ($\beta_0$) of a system is recognized as the second-order NLO response coefficient [46]. Our computations reveal that the $\beta_0$ value of $C_6Li_6$ is merely 0.5 a.u. Furthermore, $\beta_X$ and $\beta_Y$ components are negligibly small so that $\beta_0$

= $|\beta_Z|$. The dependence of $\beta_0$ of $C_6Li_6$ on the external electric field is plotted in Fig. 5. It indicates that the $\beta_0$ of $C_6Li_6$ first increases linearly and then rapidly with the increase in the OEEF values. For instance, the $\beta_0$ ranges from $2.6 \times 10^3$ a.u. (OEEF = $10 \times 10^{-4}$ a.u.) to $9.3 \times 10^3$ a.u. (OEEF = $30 \times 10^{-4}$ a.u.) and becomes as high as $3.4 \times 10^4$ a.u. for OEEF = $50 \times 10^{-4}$ a.u. Thus, the $\beta_0$ value of $C_6Li_6$ is significantly enhanced by applying an external electric field. These $\beta_0$ values, being in the order of $10^3$-$10^4$ a.u., suggest the strong NLO response of $C_6Li_6$ in the presence of an external field. Note that these $\beta_0$ values of the $C_6Li_6$ molecule are comparable to those of some superatom compounds [23] such as $(BLi_6)(BeF_3)$ and $(BLi_6)(BF_4)$ under the same OEEF. Furthermore, the $\beta_0$ value of $C_6Li_6$ for OEEF = $50 \times 10^{-4}$ a.u. is also comparable to the $\beta_0$ value of $8.5 \times 10^4$ a.u. for Li@GDY as reported by Li and Li [12]. These results also support the claim that the $\beta_0$ value of $C_6Li_6$ molecule can be further enhanced by strengthening the applied electric field.

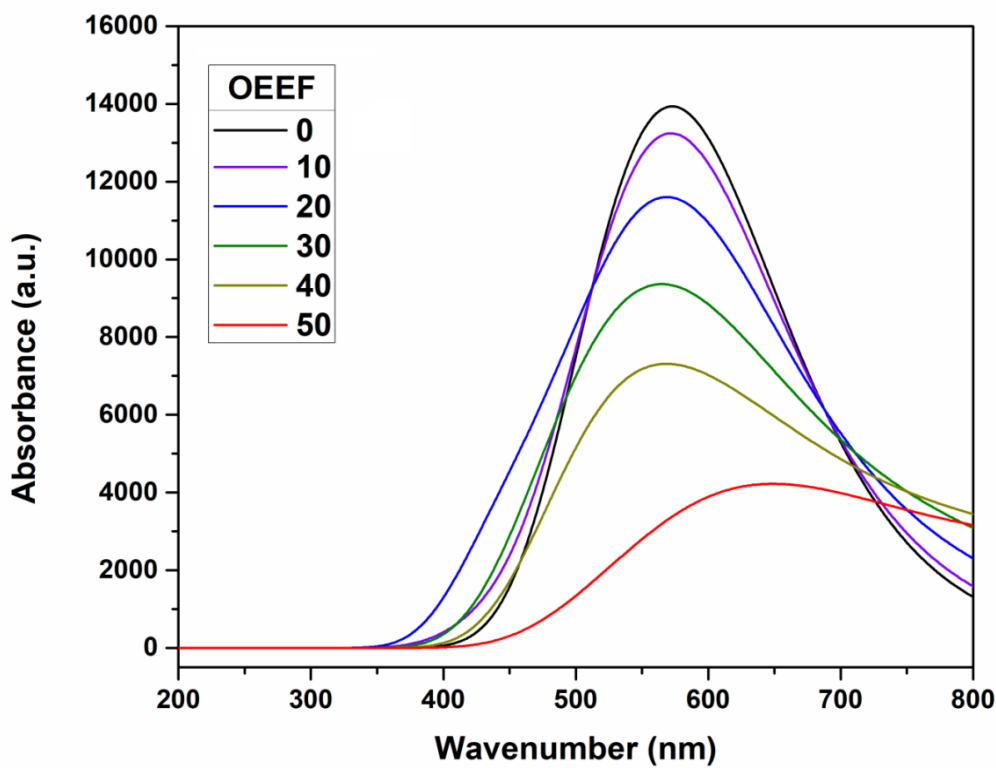

Fig. 6. TD-DFT calculated UV-vis spectra of $C_6Li_6$ molecule for various oriented external electric field (OEEF) applied (in units of $10^{-4}$ a.u.). The (maximum) absorption wavelengths are listed in Table 2.

To analyze the optical absorption of $C_6Li_6$ under the effect of OEEF, we have performed the TD-DFT//CAM-B3LYP/6-311+G(d) calculations. TD-DFT calculated UV-vis absorption spectra of $C_6Li_6$ are plotted in Fig. 6 for various OEEF values and absorption wavelength ($\lambda_{max}$) is listed in Table 2. In the absence of OEEF, the absorption peak corresponds to HOMO-1→LUMO+1 and HOMO→LUMO+2 transitions with a contribution of 40% each. This peak is observed at $\lambda_{max}$= 572.7 nm, which lies in the visible region. By increasing the OEEF, from 10 to 30 ×10$^{-4}$ a.u., $\lambda_{max}$ is only slightly shifted towards the blue region, lying in the range 569.8-560.9 nm. However, this blue shift becomes significant (> 30 nm) for OEEF = 40×10$^{-4}$ a.u., with $\lambda_{max}$ at 528.9 nm, corresponding to HOMO→LUMO+8 (42%) and HOMO→LUMO+12 (23%) transitions. For OEEF = 50×10$^{-4}$ a.u., on the contrary, there is a red-shift of 62 nm so that $\lambda_{max}$, corresponding to HOMO-1→LUMO+7 (40%) transition, is obtained at 634.8 nm. Thus, the visible transparency of $C_6Li_6$ molecule with and without field suggests its possible application in molecular electronics such as in OLEDs and optoelectronic devices. Furthermore, the significant blue/red shifts in $\lambda_{max}$ can be used to explain the enormously high $\beta_0$ value of $C_6Li_6$ under the effect of higher OEEF. For instance, the $\beta_0$ for OEEF = 50 ×10$^{-4}$ a.u. becomes twice to that for OEEF = 40 ×10$^{-4}$ a.u. According to the two-level expression, the static $\beta_0$ can be approximated [47] as follows:

$$\beta_o \propto \frac{\Delta\mu \cdot f_o}{E^3}$$

It is evident, from the above expression, that $\beta_0$ is directly proportional to the transition dipole moment ($\Delta\mu$) corresponding to $\lambda_{max}$, which is specified by the largest oscillator strength ($f_0$) value. Nevertheless, it is inversely proportional to the third power of transition energy ($\Delta E$), which appears as the most significant factor in determining the $\beta_0$ value. For OEEF = 40 ×10$^{-4}$ a.u., our calculations lead, $\Delta\mu$ =1.726 D, $f_0$ = 0.099 a.u. and $\Delta E$ = 2.344 eV.

For OEEF = 50 ×10$^{-4}$ a.u., we obtain, $\Delta\mu$ =1.122 D, $f_0$ = 0.054 a.u. and $\Delta E$ = 1.953 eV. One can readily observe that the values of $\Delta\mu$, $f_0$ as well as $\Delta E$ decrease by increasing the OEEF from 40 to 50 ×10$^{-4}$ a.u. However, the enormously high value of $\beta_0$ for OEEF = 50 ×10$^{-4}$ a.u. results due to the dominant power factor of $\Delta E$. Thus, the enhanced NLO response of the C$_6$Li$_6$ molecule with the external field advocates its possible application in nonlinear optics.

## 4. Conclusions

Our study demonstrated that it is possible to enhance the first-order mean hyperpolarizability ($\beta_0$) of C$_6$Li$_6$ by applying the OEEF. Under the OEEF, the C$_6$Li$_6$ molecule remains planar, with the slight change in C-C and C-Li bond lengths. The total energy and VIE values suggested that the stability of C$_6$Li$_6$ is enhanced to some extent by the OEEF applied. We have also noticed that the HOMO-LUMO gap of C$_6$Li$_6$ gradually decreases by increasing the OEEF, suggesting its increased conductivity, which was attributed to the shifting of the LUMO level. The dipole moment of C$_6$Li$_6$, ranging between 0.002-1.700 D, increases linearly with the increase in the OEEF. The mean polarizabilities of C$_6$Li$_6$ also increase from 241.5 a.u. to 272.2 a.u. More importantly, the $\beta_0$ value is increased from 2.5 ×10$^3$-9.2 ×10$^3$ a.u. for OEEF = 10 ×10$^{-4}$-30 ×10$^{-4}$ a.u., becoming 1.6×10$^4$ a.u. for OEEF = 40 ×10$^{-4}$ a.u. and as large as 3.4 ×10$^4$ a.u. for OEEF = 50 ×10$^{-4}$ a.u. Thus, this study provides a suitable path for enhancement of NLO response of C$_6$Li$_6$ and related systems, which might appear as suitable candidates for the design of potential NLO materials. Furthermore, we have also noticed the visible transparency of C$_6$Li$_6$ with and without OEEF, which suggests its possible applications in optoelectronics and OLEDs.

**Acknowledgement**

Dr. A. K. Srivastava acknowledges Prof. N. Misra, Department of Physics, University of Lucknow, Prof. S. N. Tiwari, Department of Physics, DDU Gorakhpur University, for helpful discussions and University Grants Commission (UGC), New Delhi, India for approving Start Up project [Grant No. 30-466/2019(BSR)].

**References**

[1] S. R. Marder, *Chem. Commun.,* 2006, 131-134.

[2] C. Zhang, Y. Song and X. Wang, *Coord. Chem. Rev.,* 2007, **251**, 111-141.

[3] D. R. Kanis, M. A. Ratner and T. J. Marks, *Chem. Rev.,* 1994, **94**, 95-242.

[4] G. de la Torre, P. Vázquez, F. Agulló-López and T. Torres, *Chem. Rev.,* 2004, **104**, 3723-3750.

[5] R.-L. Zhong, H.-L. Xu, Z.-R. Li and Z.-M. Su, *J. Phys. Chem. Lett.,* 2015, **6**, 612-619.

[6] W. Chen, Z.-R. Li, D. Wu, F.-L. Gu, X.-Y. Hao, B.-Q. Wang, R.-J. Li and C.-C. Sun, *J. Chem. Phys.,* 2004, **121**, 10489-10494.

[7] Y. Li, Z.-R. Li, D. Wu, R.-Y. Li, X.-Y. Hao and C.-C. Sun, *J. Phys. Chem. B,* 2004, **108**, 3145-3148.

[8] J. Lin, M.-H. Lee, Z.-P. Liu, C. Chen and C. J. Pickard, *Phys. Rev. B,* 1999, **60**, 13380-13389.

[9] F. Liang, L. Kang, P. Gong, Z. Lin and Y. Wu, *Chem. Mater.,* 2017, **29**, 7098-7102.

[10] M. Kalmutzki, M. Strobele, F. Wackenhut, A. J. Meixner and J. Meyer, *Angew. Chem., Int. Ed.,* 2014, **53**, 14260-14263.

[11] X. Li, *J. Mater. Chem. C,* 2018, **6**, 7576-7583.

[12] X. Li and S. Li, *J. Mater. Chem. C,* 2019, **7**, 1630-1640.

[13] S. Shaik, S. P. de Visser and D. Kumar, *J. Am. Chem. Soc.,* 2004, **126**, 11746-11749.

[14] S. Ciampi, N. Darwish, H. M. Aitken, I. Díez-Pérez and M. L. Coote, *Chem. Soc. Rev.,* 2018, **47**, 5146-5164.

[15] S. Shaik, R. Ramanan, D. Danovich and D. Mandal, *Chem. Soc. Rev.,* 2018, **47**, 5125-5145.

[16] F. Che, J. T. Gray, S. Ha, N. Kruse, S. L. Scott and J.-S. McEwen, *ACS Catal.,* 2018, **8**, 5153-5174.

[17] S. Shaik, D. Mandal and R. Ramanan, *Nat. Chem.,* 2016, **8**, 1091-1098.

[18] G. D. Harzmann, R. Frisenda, H. S. J. van der Zant and M. Mayor, *Angew. Chem., Int. Ed.,* 2015, **54**, 13425-13430.

[19] Y. W. Son, M. L. Cohen and S. G. Louie, *Nature,* 2006, **444**, 347-349.

[20] J. Wang, X. Jin, Z. Liu, G. Yu, Q. Ji, H. Wei, J. Zhang, K. Zhang, D. Li, Z. Yuan, J. Li, P. Liu, Y. Wu, Y. Wei, J. Wang, Q. Li, L. Zhang, J. Kong, S. Fan and K. Jiang, *Nat. Catal.,* 2018, **1**, 326-331.

[21] L. Yue, J. Li, S. Zhou, X. Sun, M. Schlangen, S. Shaik and H. Schwarz, *Angew. Chem., Int. Ed.,* 2017, **56**, 10219-10223.

[22] M. Liu, Y. Pang, B. Zhang, P. De Luna, O. Voznyy, J. Xu, X. Zheng, C. T. Dinh, F. Fan, C. Cao, F. P. G. de Arquer, T. S. Safaei, A. Mepham, A. Klinkova, E. Kumacheva, T. Filleter, D. Sinton, S. O. Kelley and E. H. Sargent, *Nature,* 2016, **537**, 382-386.

[23] W.-M. Sun, C.-Y. Li, J. Kang, D. Wu, Y. Li, B.-L. Ni, X.-H. Li and Zhi-Ru Li, *J. Phys. Chem. C,* 2018, **122**, 7867-7876.

[24] L. A. Shimp, C. Chung and R. J. Lagow, *Inorg. Chim. Acta,* 1978, **29**, 77-81.

[25] R. J. Baran, Jr., D. A. Hendrickson, D. A. Laude, Jr. and R. J. Lagow, *J. Org. Chem.,* 1992, **57**, 3759-3760.

[26] Y. Xie and H. F. Schaefer III, *Chem. Phys. Lett.,* 1991, **179**, 563-567.

[27] B. J. Smith, *Chem. Phys. Lett.,* 1993, **207**, 403-406.

[28] S. M. Bachrach and J. V. Miller Jr., *J. Org. Chem.,* 2002, **67**, 7389-7398.

[29] Y.-B. Wu, J.-L. Jiang, R.-W. Zhang and Z.-X. Wang, *Chem. Eur. J.,* 2010, **16**, 1271-1280.

[30] A. K. Srivastava, *Mol. Phys.,* 2018, **116**, 1642-1649.

[31] A. Vasquez-Espinal, R. Pino-Rios, P. Fuentealba, W. Orellana and W. Tiznado, *Int. J. Hydrogen Energy,* 2016, **41**, 5709-5715.

[32] S. Giri, F. Lund, A. S. Nunez and A. Toro-Labbe, *J. Phys. Chem. C,* 2013, **117**, 5544-5551.

[33] A. K. Srivastava, *Int. J. Quantum Chem.,* 2019, **119**, e25904.

[34] S. G. Raptis, M. G. Papadopoulos and A. J. Sadlej, *Phys. Chem. Chem. Phys.,* 2000, **2**, 3393-3399.

[35] A. D. Becke, *Phys. Rev. A,* 1988, **38**, 3098-3100.

[36] C. Lee, W. Yang and R. G. Parr, *Phys. Rev. B,* 1988, **37**, 785-789.

[37] M.J. Frisch, G.W. Trucks, H.B. Schlegel, G.E. Scuseria, M.A. Robb, J.R. Cheeseman, G. Scalmani, V. Barone, B. Mennucci, G.A. Petersson, H. Nakatsuji, M. Caricato, X. Li, H.P. Hratchian, A.F. Izmaylov, J. Bloino, G. Zheng, J.L. Sonnenberg, M. Hada, M. Ehara, K. Toyota, R. Fukuda, J. Hasegawa, M. Ishida, T. Nakajima, Y. Honda, O. Kitao, H. Nakai, T. Vreven, J. Montgomery, J. A., J.E. Peralta, F. Ogliaro, M. Bearpark, J.J. Heyd, E. Brothers, K.N. Kudin, V.N. Staroverov, R. Kobayashi, J. Normand, K. Raghavachari, A. Rendell, J.C. Burant, S.S. Iyengar, J. Tomasi, M. Cossi, N. Rega, J.M. Millam, M. Klene, J.E. Knox, J.B. Cross, V. Bakken, C. Adamo, J. Jaramillo, R. Gomperts, R.E. Stratmann, O. Yazyev, A.J. Austin, R. Cammi, C. Pomelli, J.W. Ochterski, R.L. Martin, K. Morokuma, V.G. Zakrzewski, G.A. Voth, P. Salvador, J.J. Dannenberg, S. Dapprich, A.D. Daniels, O. Farkas, J.B. Foresman, J.V. Ortiz, J. Cioslowski, D.J. Fox, in, Gaussian 09, Revision D. 01, Gaussian, Inc., Wallingford CT, 2009.

[38] M. D. Harmony, V. W. Laurie, R. L. Kuczkowski, R. H. Schwendeman, D. A. Ramsay, F. J. Lovas, W. J. Lafferty and A. G. Maki, *J. Phys. Chem. Ref. Data,* 1979, **8**, 619-722.

[39] G. I. Nemeth, H. L. Selzle, and E. W. Schlag, *Chem. Phys. Lett.,* 1993, **215**, 151-155.

[40] T. Yanaia, D. P. Tew and N. C. Handy, *Chem. Phys. Lett.,* 2004, **393**, 51-57.

[41] A. D. Buckingham, *Adv. Chem. Phys.,* 1967, **12**, 107-142.

[42] H. D. Cohen and C. C. J. Roothaan, *J. Chem. Phys.,* 1965, **43**, 534-539.

[43] D. Moreno, G. Martinez-Guajardo, A. Diaz-Celaya, J. M. Mercero, R. de Coss, N. Perez-Peralta and G. Merino, *Chem. Eur. J.,* 2013, **19**, 12668-12672.

[44] T. A. Koopmans, *Physica,* 1933, **91**, 104-113.

[45] P. K. Chattaraj, H. Lee, and R. G. Parr, *J. Am. Chem. Soc.,* 1991, **113**, 1855-1856.

[46] A. D. Buckhgham and B. J. Orr, *Q. Rev. Chem. Soc.,* 1967, **21**, 195-212.

[47] J. L. Oudar, *J. Chem. Phys.,* 1977, **67**, 446-457.